\documentclass[12pt]{article}
\pdfoutput=1
\usepackage{jheppub}
\usepackage[utf8]{inputenc}
\usepackage{verbatim}
\usepackage{amsmath}
\usepackage{amssymb}
\usepackage{amsthm}
\usepackage{slashed}
\usepackage{amsfonts}
\usepackage{dsfont}
\usepackage[dvipsnames]{xcolor}
\usepackage{faktor}
\usepackage{tensor}
\usepackage{physics}
\usepackage{multirow}

\newcommand{\D}{\textnormal{d}}
\newcommand{\be}{\begin{equation}}
\newcommand{\ee}{\end{equation}}
\newcommand{\bfig}{\begin{figure}\begin{center}}
\newcommand{\efig}{\end{center}\end{figure}}
\newcommand{\bi}{\begin{itemize}}
\newcommand{\ei}{\end{itemize}}

\newcommand{\wt}{\widetilde}

\theoremstyle{definition}

\newcommand{\M}{\mathcal{M}}
\newcommand{\R}{\mathbb{R}}

\newcommand{\Z}{\mathbb{Z}}
\newcommand{\id}{\mathds{1}}
\newcommand{\rw}{\rightarrow}
\newcommand{\Gtilde}{\widetilde{G}}

\newcommand{\Omegatilde}{\widetilde{\Omega}}

\newcommand{\mc}{\mathcal}
\newcommand{\so}{\mathfrak{so}}
\newcommand{\SO}{SO^+(2,1)}
\newcommand{\TSO}{\wt{SO^+}(2,1)}
\newcommand{\Q}{\mathcal{Q}}

\begin{document}
\title{Phase space of Jackiw-Teitelboim gravity with positive cosmological constant}
\author[a]{Elba Alonso-Monsalve}
\author[a]{Daniel Harlow}
\author[b]{Patrick Jefferson}
\affiliation[a]{Center for Theoretical Physics\\ Massachusetts Institute of Technology, 77 Massachusetts Avenue, Cambridge, MA 02139, USA}
\affiliation[b]{William H. Miller III Department of Physics and Astronomy\\ Johns Hopkins University, 3400 North
Charles Street, Baltimore, MD 21218, USA}
\emailAdd{elba\_am@mit.edu}
\emailAdd{harlow@mit.edu}
\emailAdd{pjeffer4@jh.edu}
\abstract{In this paper we construct the classical phase space of Jackiw-Teitelboim gravity with positive cosmological constant on spatial slices with circle topology.  This turns out to be somewhat more intricate than in the case of negative cosmological constant; this phase space has many singular points and is not even Hausdorff.  Nonetheless, it admits a group-theoretic description which is quite amenable to quantization.}
\maketitle

\section{Introduction}
One of the biggest problems in theoretical physics is how to apply quantum mechanics to the universe as a whole.  We are quite far from being able to address this problem in realistic theories of quantum gravity. For example, although string theory quite plausibly could have four-dimensional solutions with realistic physics, we do not yet have concrete examples and, in any case, a fully non-perturbative description seems far in the future.  In this paper we retreat from this lofty goal and instead study a far simpler problem: the classical phase space of the Jackiw-Teitelboim (JT) model of gravity in $1+1$ dimensions with positive cosmological constant.  This may seem like we have retreated too far, and perhaps we have, yet there are some grounds for hope:
\bi
\item One of the more confusing aspects of quantum cosmology is its unrestricted diffeomorphism invariance: in a closed universe all diffeomorphisms are gauged, so there is no gauge-invariant notion of time evolution and no boundary with respect to which we could define diffeomorphism-invariant observables (see e.g. \cite{diraclectures,DeWitt:1967yk}).  This problem is manifest in full force for JT gravity, and so by studying JT gravity we can learn about it in a (reasonably) safe environment.
\item In recent years we have learned that the path integral formulation of quantum gravity knows a surprising amount about the non-perturbative structure of quantum gravity. For example, it knows the unitary Page curve of an evaporating black hole \cite{Penington:2019npb,Almheiri:2019psf} and the exact microstate counts of BPS black holes \cite{Dabholkar:2014ema,Iliesiu:2022kny}.  In particular, this is true for the AdS version of JT gravity \cite{Saad:2019lba}. 
It is natural to hope that the dS path integral in JT gravity also has something to teach us, and calculations of such path integrals (see, e.g., \cite{Cotler:2024xzz, Fumagalli:2024msi}) often involve cutting and gluing constructions which can be and typically are interpreted in terms of the quantized phase space of the theory.
\item Another puzzling feature of quantum cosmology is the nature of the overall size of the universe.  For one thing, it is not so clear how to think about parts of the wave function where this size is small \cite{Hartle:1983ai}, and for another, the classic calculation of the cosmic microwave background fluctuations produced during inflation runs into trouble if we try to apply it to the overall size of the universe (see \cite{Maldacena:2024uhs} for a recent review). JT gravity with positive cosmological constant includes this size mode, and thus gives a laboratory where we can study this problem.
\ei
In this paper we will not resolve any of these problems; as the very first step to addressing them, constructing the classical phase space already proves more subtle than one might have expected.  Indeed, the problem has been considered before in the literature \cite{Henneaux:1985nw,Navarro-Salas:1992bwd,Levine:2022wos,Nanda:2023wne,Cotler:2019nbi}, but none of the references have identified all of the solutions that we will present here.  The full set of solutions includes eternal expanding universes, universes with big bangs and/or big crunches, and universes with arbitrary numbers of wormholes connecting arbitrary numbers of expanding de Sitter regions.

\bfig
\includegraphics[height=3cm]{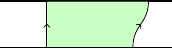}
\caption{Global information in a $dS_2$ geometry: we glue a timelike geodesic to its image under a $dS_2$-isometry in the extended $dS_2$ solution, with gluings in different conjugacy classes being physically distinct solutions (we also need to identify each gluing isometry with its inverse).  This gives one of the two phase space parameters of the theory, with the other coming from the magnitude of the dilaton.}\label{dsgluefig}
\efig
On spacetimes with no boundary the action of Jackiw-Teitelboim gravity is
\be\label{JTaction}
S=\Phi_0\int_\M d^2x \sqrt{-g}R+\int_\M d^2 x\,\sqrt{-g} \,\Phi(R-2),
\ee
where $\Phi$ is a dynamical scalar field conventionally called the dilaton and $R$ is the Ricci scalar for a dynamial metric $g_{\mu\nu}$.  We have chosen units for the cosmological constant so that the de Sitter radius is one.  The equations of motion obtained by varying this action are
\begin{equation}
\begin{split}
R-2&=0,\\
\left(\nabla_\mu\nabla_\nu+g_{\mu\nu}\right)\Phi&=0.\label{eom}
\end{split}
\end{equation}
The first equation says that there are no local gravitational degrees of freedom, so any gauge-invariant information in the metric has to do with the global structure of spacetime.  More precisely, there is one continuous degree of freedom, contained in how the geometry is glued together as we go around the cylinder (see figure \ref{dsgluefig}).  The dilaton also only has one continuous degree of freedom, which we can identify by noting that the equation of motion implies that the quantity
\be
\kappa=\nabla_\mu\Phi\nabla^\mu\Phi+\Phi^2
\ee
is constant throughout the spacetime, and thus gives another diffeormophism-invariant phase space variable.  The phase space is therefore locally two-dimensional, just as was found for the $AdS_2$ version of JT gravity in \cite{Harlow:2018tqv}, but we will see that its global structure is more complicated. 

What we will conclude is that the global structure of the phase space is\begin{small}
\begin{equation}
    \faktor{T^\ast\left(\widetilde{SO^+}(2,1)\backslash\id\right)}{\{ \sim Ad^\ast,\,\sim Inv^\ast\} } \quad \text{with}\quad Pq=P\quad\text{for all}\quad (q,P)\in T^\ast\left(\widetilde{SO^+}(2,1)\backslash\id\right),\label{ps1}
\end{equation}
\end{small}
\hspace{-.135cm}where $\widetilde{SO^+}(2,1)$ is the universal cover of the identity component of the 3-dimensional Lorentz group, $\widetilde{SO^+}(2,1)\backslash\id$ indicates this group with the identity removed, and where $q\in \widetilde{SO^+}(2,1)\backslash\id$ and $P\in T^\ast_q\left(\widetilde{SO^+}(2,1)\backslash\id\right)$. The quotient identifies elements of $T^\ast\left(\widetilde{SO^+}(2,1)\backslash\id\right)$ which are related by the codifferential of the adjoint map $Ad_h:q\mapsto hqh^{-1}$ and the codifferential of the inverse map $Inv:q\mapsto q^{-1}$, and it is also necessary to impose the constraint that the cotangent vectors $P \in T^\ast_q\left(\widetilde{SO^+}(2,1)\backslash\id\right)$ at the point $q$ are left invariant by the action of $q$ ($Pq=P$), as we will explain later.  We will also argue that this space can be rewritten simply as
\be\label{ps2}
T^*\left(\faktor{\wt{SO^+}(2,1)\backslash\id}{\sim Ad,\sim Inv}\right),
\ee
provided that sufficient care is taken in defining the tangent space at the singular points of the quotient space. The symplectic form on this phase space is given by
\be
\Omega=\delta P_i \wedge \delta Q^i,
\ee
where $Q^i$ is defined by
\be
\pi(q)=e^{Q^iT_i}
\ee
where $\pi:\widetilde{SO^+}(2,1)\to SO^+(2,1)$ is the covering map, $T_i$ are the generators of the Lie algebra of $SO^+(2,1)$, and the norm of $Q^i$ is taken to be greater than $-\pi^2$.  The expression \eqref{ps2} is fairly close to the phase space proposed for this theory back in 1992 by \cite{Navarro-Salas:1992bwd}, but they didn't have the universal cover or the quotient by $Inv$, and thus both missed solutions and treated as distinct solutions which are not distinct.  We also compute the symplectic form on this phase space using the covariant phase space formalism, confirming that locally it is two-dimensional (away from singular points).    

This paper is organized as follows. In Section \ref{sec:sols} we solve for all the classical solutions and outline their physical meaning. In Section \ref{sec:phase-space-Omega} we describe the phase space in detail and find the symplectic form by means of a gravitational calculation. Section \ref{sec:group-theory} works out the group-theoretic construction of the phase space and the symplectic form quoted above. Finally, we comment on quantization in Section \ref{sec:quant}.

\section{Classical solutions}
\label{sec:sols}
In this section we will construct the full set of solutions of the JT gravity equations of motion \eqref{eom} with the spacetime topology of a cylinder.  Our approach is to first construct the set of metrics of constant positive curvature and cylinder topology, after which we will discuss the dilaton solutions on these metrics.  

\subsection{Metric}
\label{sec:met}
\bfig
\includegraphics[width=0.9\linewidth]{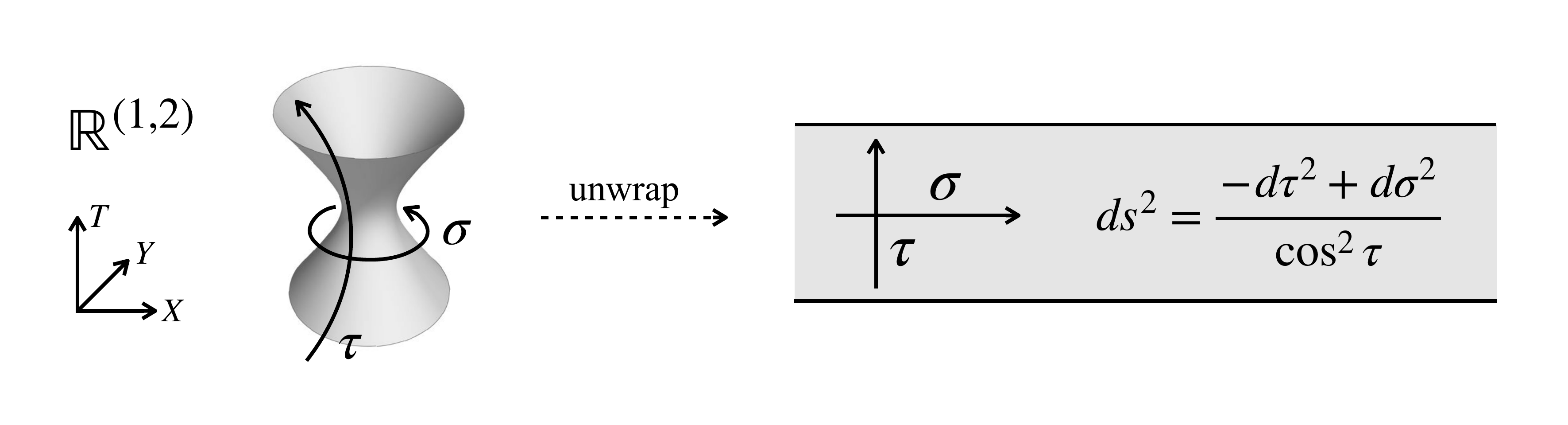}
\caption{The infinite strip, with $\tau\in(-\pi/2,\pi/2)$ and $\sigma\in\R$, is the universal cover of the de Sitter hyperboloid, (\ref{hyp}).}\label{fig:hyp}
\efig
The first thing to note is that in two spacetime dimensions the diffeomorphism class of the metric in any contractible region is entirely determined by the Ricci scalar.  Since the equations of motion set the Ricci scalar to $R=2$, this means that, locally, the metric is entirely determined up to diffeomorphism.  Since two-dimensional de Sitter space ($dS_2$) provides a metric of constant positive curvature, any other such metric must locally look like $dS_2$.  

A useful way to construct the $dS_2$ geometry is as the embedded hyperboloid
\be\label{hyp}
-T^2+X^2+Y^2=1
\ee
in $(2+1)$-dimensional Minkowski space.  We can parameterize this hyperboloid using the coordinates
\be\label{embedcoord}
T=\tan\tau,\,X=\frac{\cos\sigma}{\cos\tau},\,Y=\frac{\sin\sigma}{\cos\tau},
\ee
in terms of which the metric is
\begin{equation}
    \D s^2 = \frac{-\D\tau^2+\D\sigma^2}{\cos^2\tau}.
\label{dsglobal}
\end{equation}
In these coordinates $\tau\in \left(-\frac{\pi}{2},\frac{\pi}{2}\right)$, $\sigma$ is periodic with periodicity $2\pi$, and light moves on 45-degree lines.  This solution thus already has cylinder topology, but it will actually be useful to ``unwrap'' it to its universal cover, on which $\sigma\in \mathbb{R}$.  We will call this universal cover the infinite strip, illustrated in figure \ref{fig:hyp}.

The isometry group of $dS_2$ is inherited from the isometry group $O(2,1)$ of three-dimensional Minkowski space.  This group has four connected components, due to the possibility of temporal and spatial reflection symmetries.  In this paper we will treat spacetime inversions as global symmetries, which means that we will restrict to geometries which are both orientable and time-orientable and not identify solutions which differ by time reversal or parity (see \cite{Harlow:2023hjb} for a recent discussion of what happens when spacetime inversions are gauged). We will thus be primarily interested in the identity component 
\be
G=SO^+(2,1)
\ee
of the Lorentz group, as this is the orientation-preserving isometry group of $dS_2$.  It is important to emphasize, however, that it is \textit{not} the symmetry group of the infinite strip: in $G$ a spatial rotation by $2\pi$ (or equivalently a translation of $\sigma$ by $2\pi$) is equal to doing nothing, while on the infinite strip it isn't.  The full orientation-preserving isometry group of the infinite strip is the universal covering group of $G$, which we will call
\be
\wt{G}=\TSO.
\ee  

To construct JT solutions with cylinder topology our approach is the following: we pick an element $q\in \wt{G}$ of the isometry group of the infinite strip, and then we quotient the infinite strip by the infinite abelian subgroup of $\wt{G}$ generated by $q$.  This will give us a spacetime with fundamental group $\mathbb{Z}$ as desired (we want the cylinder), while any larger quotient would give us some more exotic topology.  The identifications which are obtained for different $q$ are not all distinct, for two reasons.  The first is that $q$ and $q^{-1}$ generate the same abelian subgroup, and thus lead to the same identification.  The second is that for any $h\in \wt{G}$ we have
\be
x\sim qx \iff hx \sim hqx\iff hx\sim hqh^{-1}hx\iff x\sim hqh^{-1}x
\label{diff}
\ee
for any point $x$ in the spacetime. Thus $q$ and $hqh^{-1}$ also lead to identical identifications.  We can therefore parameterize the set of inequivalent identifications as
\be
\TSO/\sim Ad,\sim Inv.
\ee
In other words, the set of identifications is equal to the set of conjugacy classes of $\wt{G}$, with inverse classes also identified.  $\wt{G}$ is three-dimensional as a Lie group, and generically the group orbit of $q\in \wt{G}$ is two-dimensional, so at generic points this is a one-dimensional space of identifications.  

The isometry group $\wt{G}$ is one of those unfortunate Lie groups which does not have a faithful matrix representation.  In fact, it is isomorphic to $\wt{SL}(2,\mathbb{R})$, the universal cover of $SL(2,\mathbb{R})$, which is the canonical example of such a group.  We therefore need to use more abstract methods to analyze it (afterward we summarize three useful takeaways).  By definition, the universal covering group $\wt{G}$ of a connected Lie group $G$ is equal to the set of paths in $G$ from the identity to an arbitrary element, with homotopically-equivalent paths identified.  The group multiplication is just pointwise multiplication of the paths, $g(t)h(t)=(gh)(t)$ with $t\in [0,1]$.  As a manifold the topology of $G$ is $\mathbb{S}^1\times \mathbb{R}^2$, so the topology of $\wt{G}$ is just $\mathbb{R}^3$.

\bfig
\includegraphics[width=0.45\linewidth]{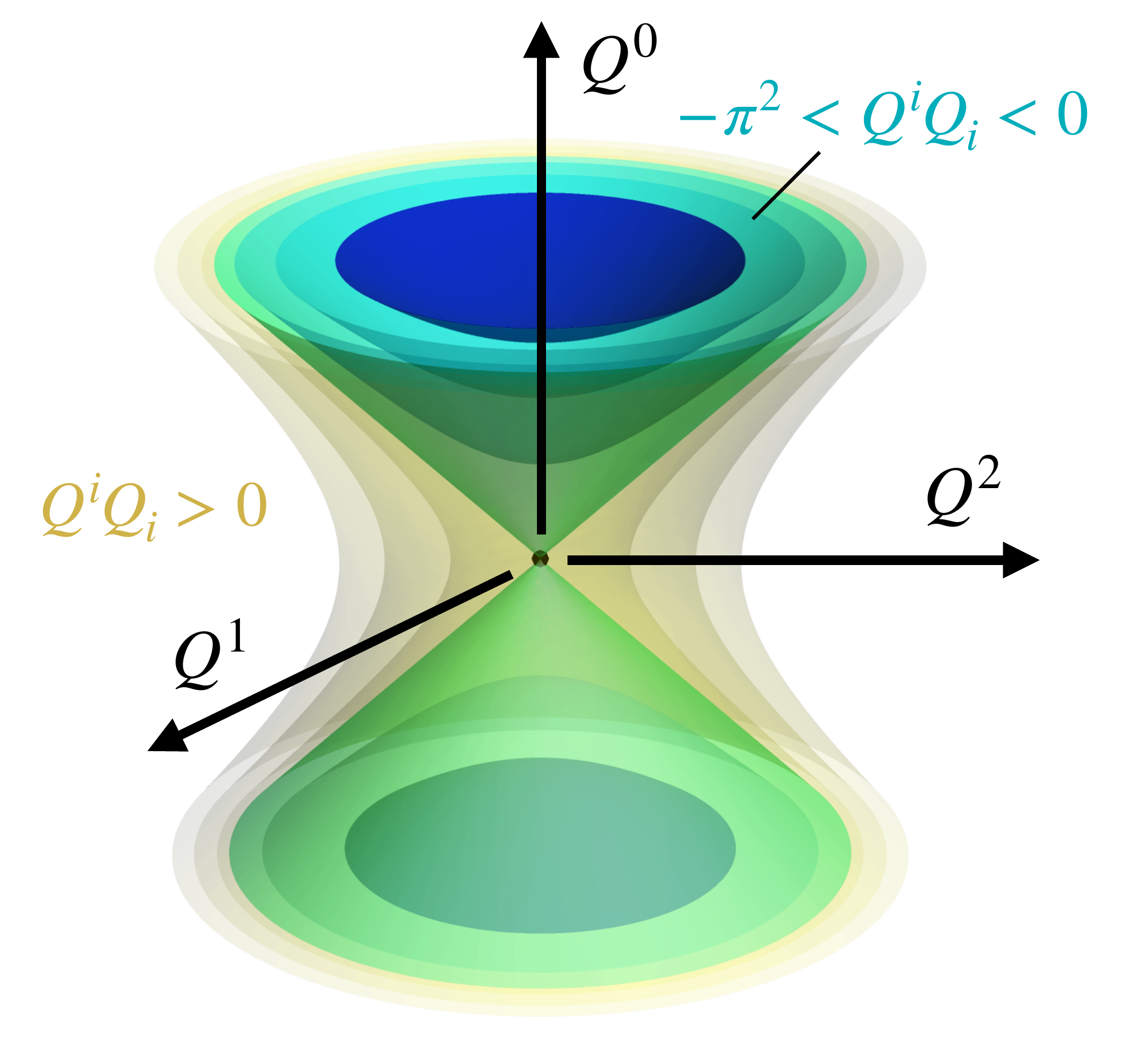}
\caption{The region $\mathcal{R}$ in three-dimensional Minkowski space which is in one-to-one correspondence with elements of $SO^+(2,1)$.  The blue points are rotations, the yellow points are boosts, and the green points are null rotations.   $\mathcal{R}$ lies between the sheets of the hyperboloid $Q^iQ_i=-\pi^2$ (dark blue), with the upper sheet included but not the lower one, since topologically points on the upper sheet are identified with their opposites on the lower sheet.  This identification illustrates the $\mathbb{S}^1\times \mathbb{R}^2$ topology of $SO^+(2,1)$.}
\label{fig:psl}
\efig
It is useful to give a more explicit description of the elements of $\wt{G}$.  We first note, as explained in appendix \ref{app:psl}, that elements of $G$ can be written using the exponential map as
\be
q=e^{\Q},
\ee
with
\be
\Q=Q^iT_i
\ee
being an element of the Lie algebra of $G$.  We can take the generators explicitly to be given by
\be
T_0=\begin{pmatrix}0&0&0\\0&0&-1\\0&1&0\end{pmatrix} \qquad T_1=\begin{pmatrix}0&0&-1\\0&0&0\\-1&0&0\end{pmatrix} \qquad
T_2=\begin{pmatrix}0&1&0\\1&0&0\\0&0&0\end{pmatrix},
\ee
and we show in appendix \ref{app:psl} that each element of $G$ is represented exactly once if we take $Q^i$ to live in the region
\be\label{Rdef}
\mathcal{R}=\{Q^i\:|\:Q_iQ^i\geq -\pi^2, Q^0>0 \:\,\mathrm{if}\:\, Q_iQ^i=-\pi^2\}.
\ee
Here
\be
Q_i Q^i=-(Q^0)^2+(Q^1)^2+(Q^2)^2\geq -\pi^2.
\label{R}
\ee
See figure \ref{fig:psl} for a graphical representation of $\mathcal{R}$. We can now describe the universal cover $\wt{G}$ more explicitly as follows.  We first define a \textbf{fundamental domain} $F\subset \wt{G}$, which consists of those homotopy classes of paths in $G$ which have a representative $Q^i(t)$ that lives entirely in the region $\mathcal{R}$.  We emphasize that $F$ is \textit{not} a subgroup of $\wt{G}$, since multiplication of elements of $F$ can easily take us out of it.  
We then observe that $\wt{G}$ has a discrete central subgroup isomorphic to $\mathbb{Z}$ consisting of homotopy classes of paths which start and end at the identity in $G$ and circle around the $\mathbb{S}^1$ some integer number of times.  A generator of this group is the operation which translates $\sigma$ by $2\pi$ to the right in the infinite strip, which corresponds to a path that circles once around the $\mathbb{S}^1$ in $G$, and we will refer to this generator as $T$.   We then have the following three useful facts:
\bi
\item[(1)] Every element $q\in\wt{G}$ can be uniquely written as the product of an element $q_0\in F$ and a central element $T^n$: 
\be
\label{uniq}
q=q_0 T^n.
\ee
This is clear from the topology of $\wt{G}$: $T^n$ just translates the fundamental domain by $2\pi n$.  By the results of appendix \ref{app:psl} we therefore can parameterize elements of $\wt{G}$ by pairs $(Q^i,n)$ where $Q^i\in \mathcal{R}$ and $n\in \mathbb{Z}$.
\item[(2)]  Conjugating a Lie algebra element $\Q$ by an element of $G$ just rotates/boosts the vector $Q^i$ by that transformation:
\be
e^{\mathcal{H}}\Q e^{-\mathcal{H}}=(e^{\mathcal{H}})^i_{\phantom{i}j}Q^jT_i.
\ee
This follows from the adjoint action of $G$ on its Lie algebra.
\item[(3)] If $q,h\in \wt{G}$ are in the fundamental domain $F$, then $hqh^{-1}$ is also in $F$.  The proof of this is that we can write $q=e^{t\Q}$ and $h=e^{t\mathcal{H}}$ with $\Q,\mathcal{H}\in \mathcal{R}$, and then the path
\be
e^{t\mathcal{H}}e^{t\Q}e^{-t\mathcal{H}}
\ee
lives entirely in $\mathcal{R}$, since it just gradually scales up the length of $\Q$ and the conjugation just boosts/rotates it around within $\mathcal{R}$.
\ei
\bfig
\includegraphics[width=0.7\linewidth]{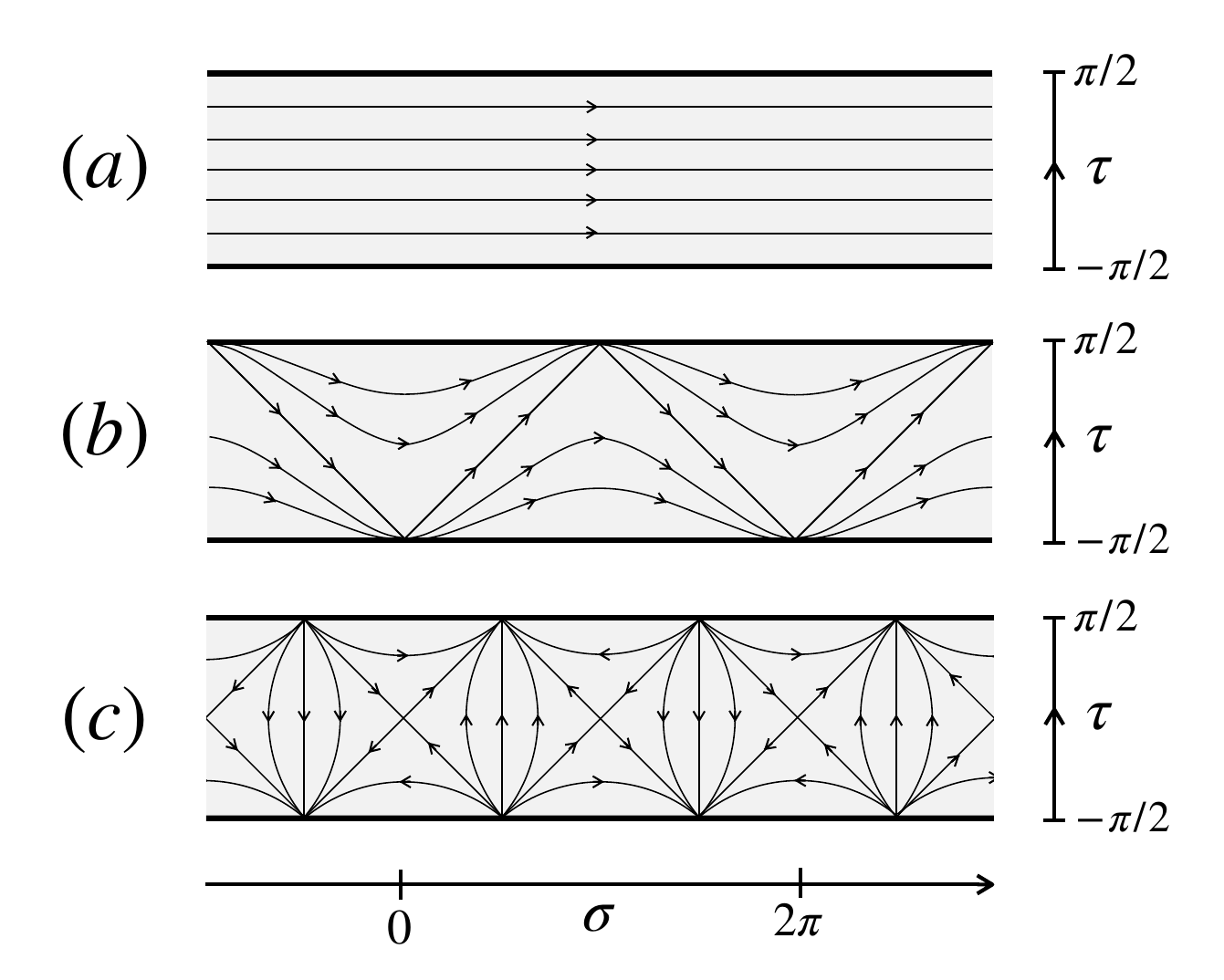}
\caption{Integral curves of Killing vector fields (KVFs) corresponding to the three nontrivial types of conjugacy classes of $\SO$. (a) rotation, with global-coordinate components $(0,1)^\mu$; (b) null rotation, with components $(\cos\tau\sin\sigma,1+\sin\tau\cos\sigma)^\mu$; (c) boost, with components $(\cos\tau\sin\sigma,\sin\tau\cos\sigma)^\mu$. 
Motion along these vector fields corresponds to elements $(Q^i,n)$ of $\Gtilde$ where $Q^i$ is (a) a timelike vector proportional to $(1,0,0)$, (b) a null vector proportional to $(1,-1,0)$, and (c) a spacelike vector proportional to $(0,-1,0)$. Note that all KVFs are invariant under the central $2\pi$ translations in the $\sigma$ direction.}\label{fig:kvfs}
\efig
Taken together these facts allow us to characterize the conjugacy classes of $\wt{G}$: given $q=q_0T^n$ and $h=h_0T^m$ in $\wt{G}$, we have
\be
hqh^{-1}=h_0q_0h_0^{-1}T^n.
\ee
The conjugacy classes of $\wt{G}$ are thus labeled by conjugacy classes of $G$ together with an integer $n$ that tells us how many times we translate $\sigma$ by $2\pi$.

\bfig
\includegraphics[width=0.63\linewidth]{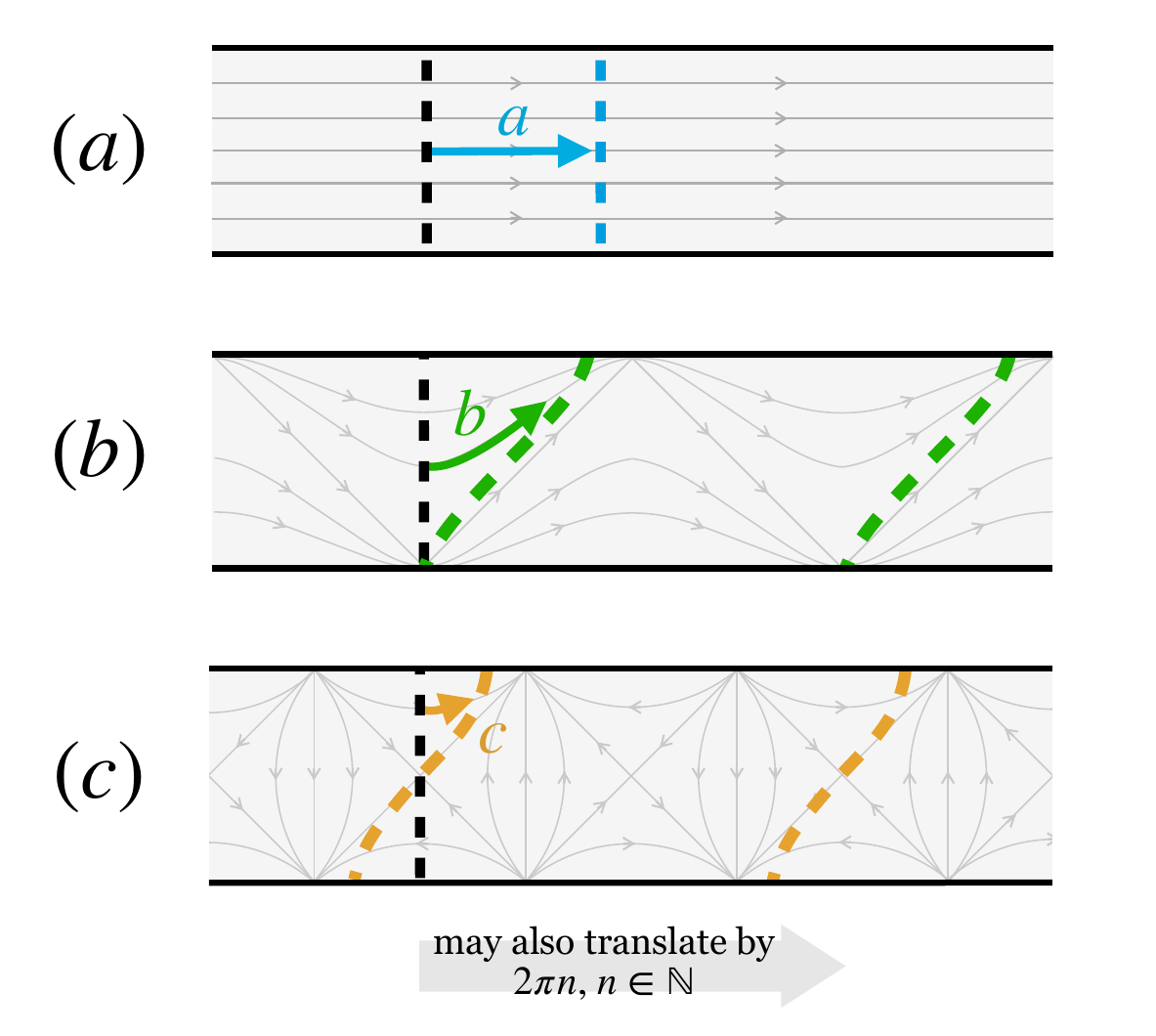}
\caption{Representative actions on the infinite strip by elements of the various conjugacy classes of $\wt{G}$.  The colorful dashed line is the image of the black dashed line $(\sigma=0)$ under the isometry. The vectors of Lie-algebra charges that generate each transformation (aside from $2\pi n$ translations) are (a) $Q^i=(a,0,0)$, timelike; (b) $Q^i=(b,-b,0)$, null; and (c) $Q^i=(0,-c,0)$, spacelike.}
\label{fig:quotients}
\efig
Let's now try to make this discussion more intuitive.  The conjugacy classes of $\SO$ consist of the trivial identity class, the class of rotations by some angle $\theta$ in some Lorentz frame, the class of boosts of rapidity $\eta$ in some Lorentz frame, and the class of ``null rotations,'' which in some Lorentz frame are generated by $T_0 + T_1$.  In figure \ref{fig:kvfs} we show integral curves in the infinite strip for Killing vector fields which generate examples of the three nontrivial classes.  Let's now consider the cylinder spacetimes which result from these classes of quotients.  Since $q$ and $q^{-1}$ give identical quotients, we can always take $n\geq 0$.  The actions of these isometries on the timelike geodesic $\sigma=0$ in the infinite strip are shown in figure \ref{fig:quotients}.   For $n\geq1$, the quotients by these actions all give a fundamental domain which heuristically resembles the gluing in figure \ref{dsgluefig}, with the fundamental domain contained between the $\sigma=0$ curve and its isometric image.\footnote{We will henceforth refer to quotients of the infinite strip as ``gluings'' for simplicity.} The case in Figure \ref{fig:quotients}(c) is akin to the timeshift described in \cite{Chen:2024rpx}; an achronal geodesic acquires a boost $|Q|$ as it wraps around the cylinder.  More nontrivial are the quotient spacetimes with $n=0$; these are shown in figure \ref{n0fig}. 
\bfig
\includegraphics[width=0.55\linewidth]{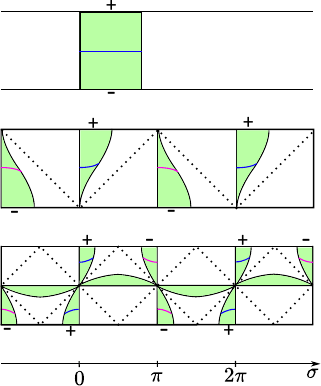}
\caption{Fundamental domains for quotients by elements of $\wt{G}$ with $n=0$ in the timelike, null, and spacelike cases.  We view the regions with blue Cauchy slices as valid solutions of the theory, while we discard the ones with pink Cauchy slices based on our rule that a solution should contain a $dS_2$ boundary where $\Phi\to +\infty$.  The sign of the boundary dilaton divergence is indicated in the figure.}\label{n0fig}
\efig
In the timelike case we get a cylinder, as hoped; in the null case we get an infinite number of cylinders, which alternate between expanding and contracting universes; and in the spacelike case we get madness: an infinite set of double cones, each with an expanding and a contracting part beginning/ending in a singularity, as well as an infinite set of ``totally vicious'' regions, where every point lies on a closed timelike curve.\footnote{This spacetime is a dS version of the ``Misner spacetime'', which is a quotient of two-dimensional Minkowski space by a finite boost \cite{Hawking:1973uf}.}  The singular points connecting the various regions are not even Hausdorff!  In the null and spacelike cases, however, we can still pick regions of the spacetime which are globally hyperbolic and have cylinder topology; Cauchy slices for these regions are shown in figure \ref{n0fig}.  The principle we will adopt is that we will view as physical any of these regions which includes a piece of $dS_2$ infinity, which, as we will see in the next section, includes a rule that the dilaton should go to $+\infty$ (and not $-\infty$) at the asymptotic past or future.  The indicated signs in figure \ref{n0fig} anticipate the signs of the dilaton divergence that we will find in the next subsection. With this rule, the timelike picture above corresponds to one expanding solution, the null figure corresponds to one expanding solution, and the spacelike figure corresponds to \textit{two} solutions: an expanding solution and a contracting solution.  The signs can be flipped by flipping the sign of the dilaton solution; in the timelike and null cases this gives us a new solution, while in the spacelike case it doesn't give us anything new.  

There is one further issue at $n=0$: what happens when we take $q$ to be the identity?  We can either try to interpret this as no gluing at all (in which case the spacetime isn't a cylinder) or as the limit of a very dense gluing, in which case space has zero volume (and the spacetime again isn't a cylinder).  Our approach will be to just exclude the identity gluing from the phase space.

\subsection{Dilaton}
\label{sec:dil}
We now need to solve the dilaton equation of motion. The full set of solutions has the form
\be
\Phi=V_i X^i=V_0\tan \tau+V_1\frac{\cos\sigma}{\cos\tau}+V_2\frac{\sin\sigma}{\cos \tau}\label{Phi}
\ee
with $X^i=(T,X,Y)$ as in (\ref{embedcoord}). These solutions can again be classified by whether the vector $V^i$ is timelike, null, or spacelike.  The lines of constant dilaton for the three kinds of solutions are closely related to the integral curves of the isometries shown in figure \ref{fig:kvfs}.  We can describe these solutions more concretely by noting that when $V^i$ is timelike we can boost it (by boosting $X^i$) so that $V_i=(V_0,0,0)$, in which case the dilaton solution and metric in global coordinates are given by
\begin{equation}
\begin{split}
ds^2&=\frac{-d\tau^2+d\sigma^2}{\cos^2\tau}\\
\Phi&=V_0\tan \tau,
\end{split}
\end{equation}
when $V^i$ is null we can rotate it so that $V_i=(V_0,V_0,0)$, in which case we can write the solution in ``flat-slicing'' coordinates 
\begin{equation}
\begin{split}
T&=\sinh \omega+\frac{1}{2}e^{\omega}x^2\\\nonumber
X&=\cosh \omega-\frac{1}{2}e^\omega x^2\\
Y&=xe^{\omega}
\end{split}
\end{equation}
as
\begin{equation}
\begin{split}
ds^2&=-d\omega^2+e^{2\omega}dx^2\\
\Phi&=V_0 e^{\omega},
\end{split}
\end{equation}
and when $V^i$ is spacelike we can rotate it so that $V_i=(0,V_1,0)$, in which case we can write the solution in ``static patch'' coordinates
\begin{equation}
\begin{split}
T&=\sqrt{1-y^2}\sinh t\\\nonumber
X&=-y\\
Y&=\sqrt{1-y^2}\cosh t
\end{split}
\end{equation}
to get
\begin{equation}
\begin{split}
ds^2&=-(1-y^2)dt^2+\frac{dy^2}{1-y^2}\\
\Phi&=-V_1y.
\end{split}
\end{equation}
The Killing vectors illustrated in figure \ref{fig:kvfs} are precisely $\partial_\sigma$, $\partial_x$, and $\partial_t$ in these coordinates, so their integral curves are the lines of constant $\Phi$.

Having understood the dilaton solutions on the infinite strip, we now want to understand when they are compatible with the spacetime quotient that turns the infinite strip into a cylinder.  The necessary and sufficient condition is quite simple to state: in order for the dilaton to be smooth on the spacetime created by quotienting the infinite strip by $q\in\wt{G}$ (or, more accurately, by the subgroup it generates), the dilaton solution needs to be invariant under $q$:
\be
\Phi(qx)=\Phi(x).
\ee
In terms of the vector $V^i$, the condition we need is that
\be
\pi(q)V=V
\ee
where 
\be
\pi:\wt{G}\to G,
\ee
is the covering map.  From now on we will just write this equation equivalently as
\be
qV=V,
\label{stab}
\ee
keeping in mind that the (non-faithful) representation of $\Gtilde$ on vectors $V$ is isomorphic to the vector representation of $G$, as the central elements of $\Gtilde$ act trivially on $V$.\footnote{The fact that the central elements of $\Gtilde = \widetilde{SO^+}(2,1)$ act trivially on $V \in \mathbb R^{1,2}$ can be seen by noting that the vector representation of $SO^+(2,1)$ is isomorphic to the adjoint representation, which is in turn isomorphic to the adjoint representation of $\widetilde{SO^+}(2,1)$, as their Lie algebras are identical.}  In terms of the decomposition $q=q_0T^n$ with $q_0=e^\Q$, this condition simply requires that
\be
Q^i\propto V^i.
\ee
Note that the coefficient of proportionality can be zero if $q_0$ is the identity.

In the previous section we saw that a gluing $q$ and a gluing $hqh^{-1}$ yield the same spacetime, evident upon a diffeomorphism $x\mapsto hx$ with $h\in\Gtilde$, as shown in (\ref{diff}). Such a diffeomorphism transforms $\Phi(x)=V_iX^i$ into $\Phi(hx)=V'_iX^i$, with the vector $(V')^i$ satisfying $V'=hV$. Therefore we should view $(q,V)$ and $(hqh^{-1},hV)$ as equivalent solutions of JT gravity. Note that this transformation preserves (\ref{stab}).

Finally, as a physical comment, purely from the point of view of the JT model it is not so clear what we are to make of the difference between pieces of the $dS_2$ boundary where the dilaton diverges to $+\infty$ and pieces where it diverges to $-\infty$.  Inspired by experience with $AdS_2$, the philosophy we will take is that the $+\infty$ boundaries are genuine $dS_2$ boundaries, while the $-\infty$ boundaries should be interpreted as curvature singularities.  This is motivated by the idea that the dilaton is really measuring the size of some extra dimension.\footnote{See, e.g., \cite{Maldacena:2019cbz}, where JT gravity on $dS_2$ is shown to arise as a limit of the 4D extremal Nariai black hole solution.}  We will therefore dismiss solutions which do not have any piece of $dS_2$ boundary (with $\Phi=+\infty$) in our discussion of solutions with $n=0$ in the previous subsection.  From a purely gravitational point of view there does not seem to be an intrinsic reason to do this, but we will see that this rule makes the phase space structure nicer and in particular more amenable to quantization.  

\section{Gravitational description}
\label{sec:phase-space-Omega}

In the covariant approach to Hamiltonian mechanics, the phase space of the theory is defined to be the set of solutions of the equations of motion modulo gauge transformations.  One then follows a standard sequence of steps to construct a symplectic form on this phase space and generators for any non-gauged symmetries. The symplectic form is a closed nondegenerate two-form on phase space which can be used to convert any function on phase space into a vector field, whose integral curves then correspond to the evolution generated by that function.  See \cite{Harlow:2019yfa} for a recent review and more references.  In this section we will give a first characterization of the $dS_2$ phase space, and then use the covariant phase space formalism to endow it with a symplectic form.

\subsection{Phase space}
In the previous section we constructed the set of solutions of JT gravity with positive cosmological constant on a Lorentzian cylinder, subject to the requirement that each solution contain some piece of $dS_2$ infinity (which we defined to have  $\Phi\rw +\infty$).  In all but two cases, such solutions are labeled by a group element $q\in \wt{G}$ and a vector $V^i$ in the embedding space, subject to the restriction
\be\label{Vinv}
qV=V
\ee
and the identifications 
\begin{equation}
\begin{split}
(q,V)&\sim (hqh^{-1},hV)\\
(q,V)&\sim (q^{-1},V).
\label{ids}
\end{split}
\end{equation}
The first case where this parametrization does not work is when $q$ is the identity; as we already discussed, this geometry is not a cylinder, so we should exclude it from the phase space.  The second case is when $n=0$ and $Q^i$ is spacelike, in which case we saw in figure \ref{n0fig} that there are two distinct solutions (the expanding and contracting parts of the double cone) but this construction only counts one of them.  For now we will just keep this discrepancy in mind; in the following section we will propose a revised parametrization which fixes it and also resolves an issue with the symplectic form that we will encounter later in this section.

\bfig
\includegraphics[width=0.9\linewidth]{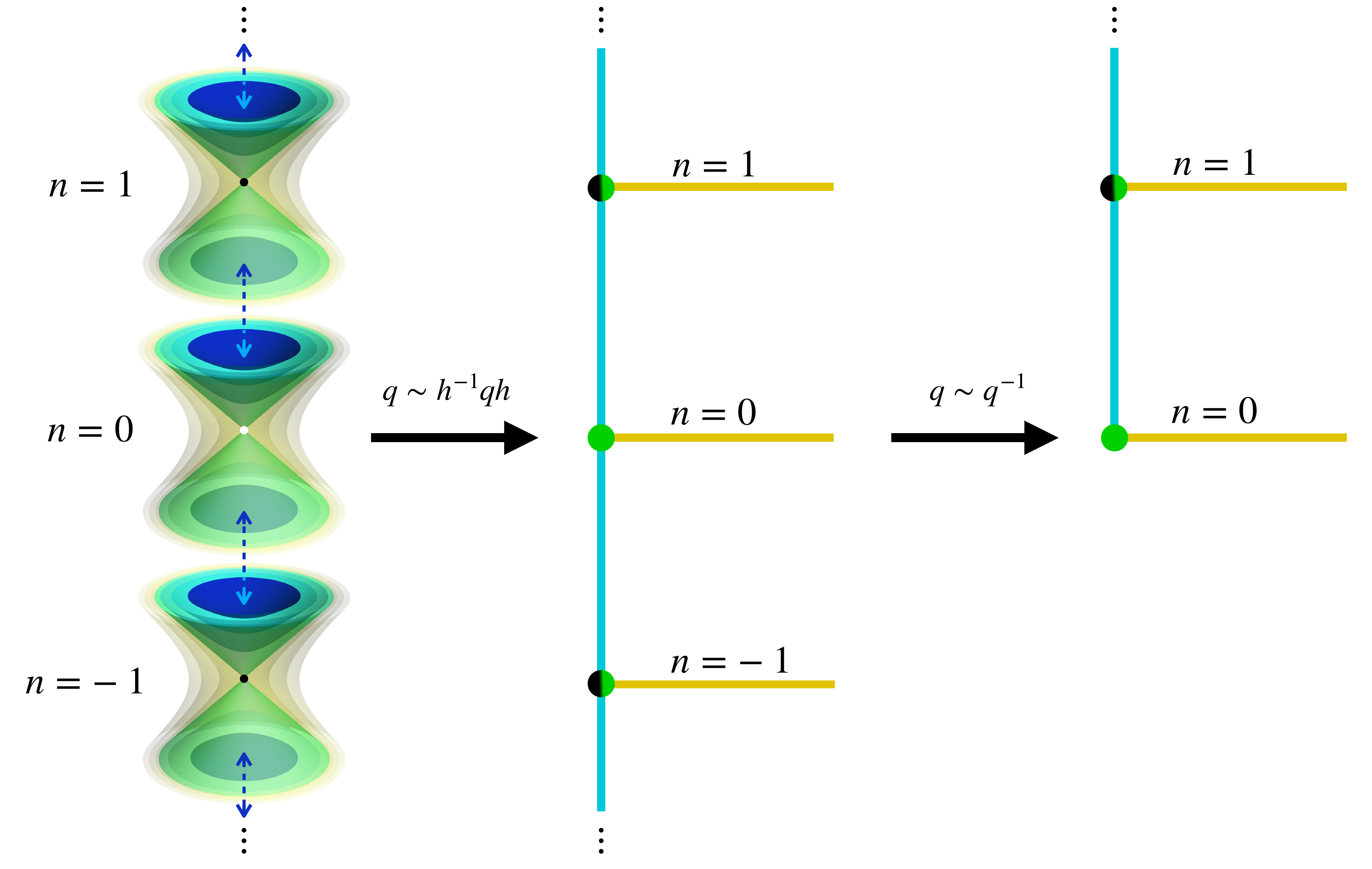}
\caption{The set of gauge-inequivalent gluings to turn the infinite strip into the cylinder, labeled by conjugacy classes of $\wt{G}$ with inverse classes identified.  The yellow spacelike classes and the blue timelike classes are locally one-dimensional, and they meet at singular points where null gluings (green) and central elements of $\wt{G}$ (black) sit on top of each other. Note that we have removed the identity element.}\label{fig:base}
\efig
This phase space has a quotient topology, which it inherits from the parent space $\mathbb{R}^6$.  This topology is somewhat weird, however.   Let's first consider the topology of the set of gauge-inequivalent gluings.  This is $\wt{G}$ quotiented by the adjoint and inverse actions.  The basic source of trouble is that when $Q^i$ is null we can act on it with a boost to send it arbitrarily close to the origin.  This causes the quotient space to not be Hausdorff, since for any open neighborhood $U_{null}$ of a null vector $Q^i$ and any open neighborhood $U_0$ of the origin, there is a boost of $U_{null}$ which intersects $U_0$.  In the quotient space, the null gluings are therefore sitting ``right on top'' of the central elements of $\wt{G}$.  When we continuously deform a spacelike vector into a timelike vector by way of a null vector, we therefore also pass through one of these singular points in the quotient.  We illustrate this singular space of gluings in figure \ref{fig:base}.

\bfig
\includegraphics[width=0.5\linewidth]{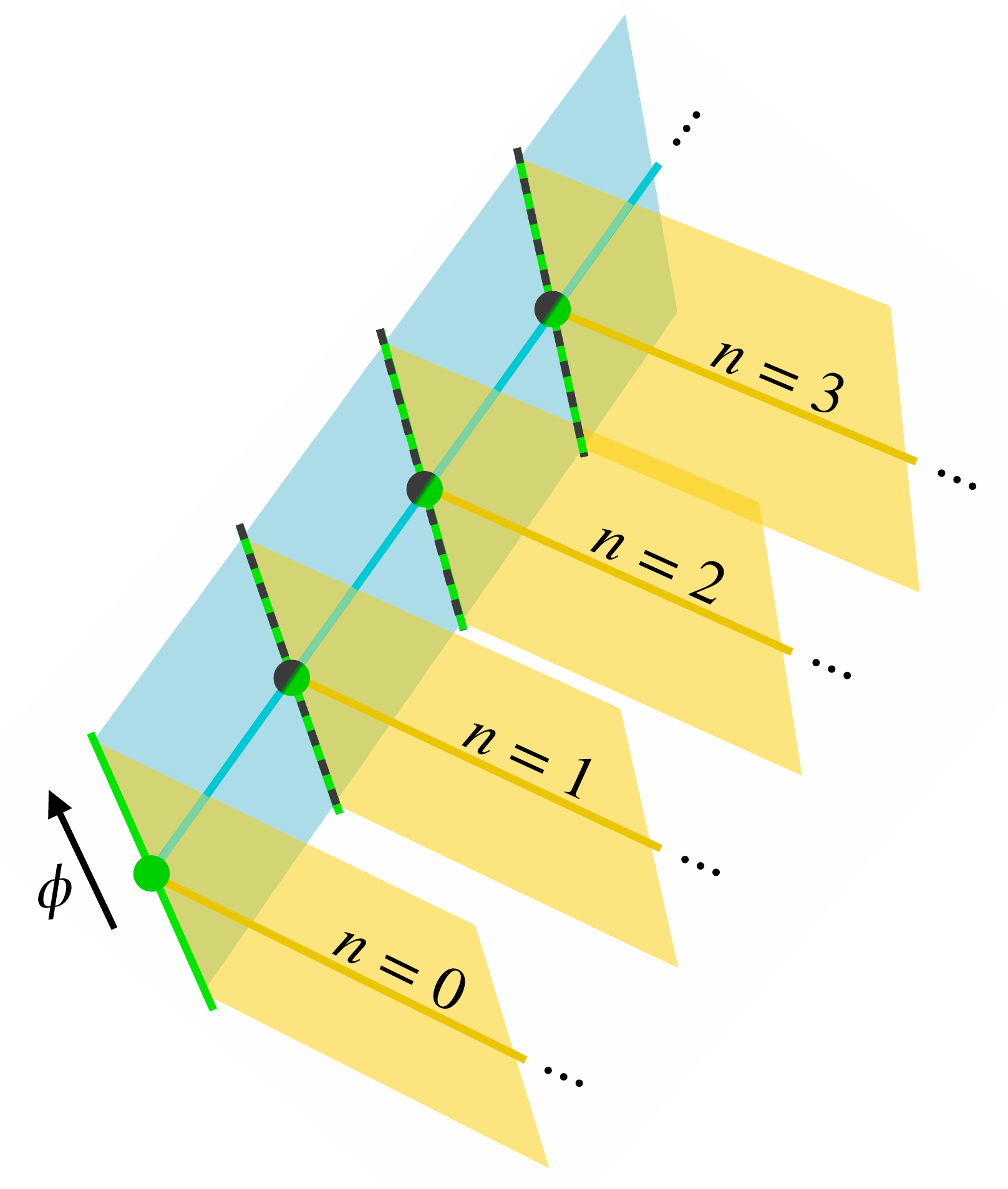}
\caption{A heuristic picture of the phase space, with dilaton fibers on top of each solution away from the singular points.  The null solutions at the singular points also have dilaton fibers on top, while the fibers for the central elements (which are on top of the null solutions due to the non-Hausdorff topology) are more bizarre.}\label{fig:phase-space}
\efig
We can now incorporate the dilaton in a simple way: away from the central elements of $\wt{G}$, the restriction \eqref{Vinv} tells us that we must have 
\be
V^i=\phi\, Q^i
\ee
for some $\phi\in \mathbb{R}$, so the dilaton simply adds a one-dimensional fiber on top of the (singular) base space we have just discussed.  At the central elements, on the other hand (excluding the identity, which we have removed), any dilaton is allowed, but we should identify vectors $V$ which differ by Lorentz transformations.  There are thus \textit{two} $\mathbb{R}$-fibers sitting on top of each central $q$: one for timelike $V$ and one for spacelike $V$, as well as two points which keep track of whether the null dilaton is future- or past-pointing. See figure \ref{fig:phase-space} for a heuristic representation of the full phase space.

\subsection{Symplectic form}
\label{sec:Omega}
In the covariant phase space formalism, the symplectic form of a Lagrangian system with action
\begin{equation}
    S=\int_{\M}L+\int_{\partial\M}\ell,
\end{equation}
with $\M$ a $d$-manifold, $L$ a $d$-form, and $\ell$ a $(d-1)$-form is constructed in the following way \cite{Harlow:2019yfa}.  The variation of the Lagrangian always has the form
\be
\delta L=E_a\delta \phi^a+\D\Theta,
\ee
where $E_a$ are the equations of motion, $\phi^a$ are the dynamical fields and $\Theta$ is a $(d-1)$-form on spacetime which is a one-form on the configuration space $\mathcal{C}$ of field histories obeying the spatial boundary conditions.  For example, for a particle moving in a potential we have $\Theta=\dot{x}\delta x$.  In this notation $\delta$ indicates the exterior derivative on $\mathcal{C}$.  In order for the theory to have a good variational principle, the action should be stationary at solutions of the equations of motion, up to boundary terms in the future and past, which means that we must have
\be
(\Theta+\delta\ell)\vert_{\Gamma}=\D C
\ee
on the spatial boundary $\Gamma$ for some scalar function $C$. We then define the ``pre-symplectic form''
\begin{equation}
    \Omegatilde=\delta\left( \int_{\Sigma}\Theta-\int_{\partial\Sigma}C\right)\Big|_{\widetilde{P}},
    \label{Omega}
\end{equation}
where $\wt{P}\subset \mathcal{C}$ is the ``pre-phase space" consisting of histories which solve the equations of motion. By construction, the pre-symplectic form (\ref{Omega}) has the property that it is independent of the Cauchy slice $\Sigma$ on which we evaluate it. On the other hand, the pre-symplectic form $\Omegatilde$ can be degenerate, and in fact its zero modes correspond precisely to continuous gauge transformations in phase space. We must quotient to remove these and render it invertible, thereby obtaining the true symplectic form $\Omega$. We now apply this construction to JT gravity with positive cosmological constant.

For the action (\ref{JTaction}) we have \cite{Harlow:2019yfa}
\begin{equation}
\begin{split}
\Theta&=\theta\cdot \epsilon\\
\theta^{\mu}&\equiv(\Phi_0+\Phi)(g^{\mu\alpha}\nabla^{\beta}-g^{\alpha\beta}\nabla^{\mu})\delta g_{\alpha\beta}+(\nabla^{\mu}\Phi g^{\alpha\beta}-\nabla^{\alpha}\Phi g^{\mu\beta})\delta g_{\alpha\beta}.
    \label{thetamu}
\end{split}
\end{equation}
Here $\epsilon$ is the spacetime volume form and $\theta\cdot \epsilon$ means inserting the vector $\theta$ into the first argument of $\epsilon$ to get a one-form.  From now on, we set $\Phi_0=0$, as  we are studying the theory classically and this term is topological, and therefore has no classical effects. We would now like to use this expression to evaluate the symplectic form of JT gravity on a circular Cauchy slice. However, there is a technical problem we need to solve: the covariant phase space formalism just described works in situations where the ranges of the spacetime coordinates do not depend on the dynamical fields.  For us the global coordinates $\tau$ and $\sigma$ which we have used to locate everything have the property that their ranges are different depending on which quotient isometry we use.\footnote{A similar problem arises in $AdS_2$ JT gravity, and we will adapt to our purposes the same method which was used there \cite{Harlow:2019yfa}.}  For each solution we therefore need to introduce a diffeomorphism whose inverse maps the global $(\tau,\sigma)$ coordinates to some new coordinates $(\hat{\tau},\hat{\sigma})$ whose ranges are the same for all solutions in an open neighborhood of the one at which we are evaluating the symplectic form.  

The details of the map between hatted and unhatted coordinates are different in the case where $n=0$ and $Q^i$ is spacelike, due to the fixed point of the quotient map, so we will first consider the other cases, where there is no fixed point. As long as there is no fixed point, we can choose the $(\hat{\tau},\hat{\sigma})$ coordinates so that their global range is simply that of a cylinder with $\hat{\sigma}\in [0,2\pi)$. We can also choose the map between the hatted and unhatted coordinates so that the surface $\sigma=0$ coincides with the surface $\hat{\sigma}=0$.  To do this we choose a diffeomorphism
\begin{equation}
\begin{split}
\tau&=f^\tau(\hat{\tau},\hat{\sigma}),\\
\sigma&=f^\sigma(\hat{\tau},\hat{\sigma}).
\end{split}
\end{equation}
with the property that it does nothing near $\hat{\sigma}=0$, while near $\hat{\sigma}=2\pi$ it acts as a central translation by $-2\pi$ followed by the gluing isometry $q$.  More concretely, near $\hat{\sigma}=2\pi$ we want
\begin{equation}
\begin{split}
    f^\tau(\hat\tau,\hat\sigma) &= \arctan\left[ \left(e^{\mc{Q}}\cdot \hat{X}\right)^0 \right], \\
    f^\sigma(\hat\tau,\hat\sigma) &= \arctan\left[ \frac{\left(e^{\mc{Q}}\cdot \hat{X}\right)^2}{\left(e^{\mc{Q}}\cdot \hat{X}\right)^1} \right]+2\pi (n-1),
\end{split}
\end{equation}
and
\begin{equation}
\begin{split}
    (f^{-1})^{\hat\tau}(\tau,\sigma) &= \arctan\left[ \left(e^{-\mc{Q}}\cdot X\right)^0 \right],\\
    (f^{-1})^{\hat\sigma}(\tau,\sigma) &= \arctan\left[ \frac{\left(e^{-\mc{Q}}\cdot X\right)^2}{\left(e^{-\mc{Q}}\cdot X\right)^1} \right]-2\pi (n-1),
\end{split}
\end{equation}
where
\begin{equation}
\begin{split}
    X^i&=\left(\tan \tau,\frac{\cos\sigma}{\cos \tau},\frac{\sin \sigma}{\cos \tau}\right),\\
    \hat{X}^i&=\left(\tan \hat{\tau},\frac{\cos\hat{\sigma}}{\cos \hat{\tau}},\frac{\sin \hat{\sigma}}{\cos \hat{\tau}}\right),
\end{split}
\end{equation}
and $(Q^i,n)$ is the gluing identification.  We emphasize that $f$ is \textit{not} an isometry; it continuously varies from doing nothing near $\hat{\sigma}=0$ to being an isometry near $\hat{\sigma}=2\pi$.  See figure \ref{dsglue2fig} for an illustration.  
\bfig
\includegraphics[height=2.5cm]{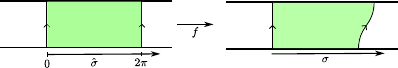}
\caption{Mapping the standard cylinder to our glued cylinder using a (non-isometric) diffeomorphism $f$. }\label{dsglue2fig}
\efig
If we do a variation in the space of solutions, this causes a variation in $f$, which we can package into an infinitesimal diffeomorphism \cite{Harlow:2019yfa}
\begin{equation}
    \xi^{\hat\alpha}(\hat\tau,\hat\sigma) = \pdv{(f^{-1})^{\hat\alpha}}{\sigma^\alpha}\Big\rvert_{f(\hat\tau,\hat{\sigma})}\delta f^\alpha (\hat\tau,\hat\sigma).
    \label{qvec}
\end{equation}
In particular, this means that, on shell, we can view the metric variation in $\Theta$ as a pure diffeomorphism:
\be
\delta g_{\mu\nu}=\mathcal{L}_\xi g_{\mu\nu}.
\ee
The machinery of the covariant phase space formalism then tells us that we will have
\be
\Theta|_{\mathcal{P}}=X_\xi\cdot \Theta=J_\xi+\xi\cdot L=dQ_\xi,
\ee
where the symbols in the intermediate step are defined in \cite{Harlow:2019yfa} and in the final result $Q_\xi$ is the ``Noether charge''
\begin{equation}
\begin{split}
Q_\xi&=-\Phi\star d\xi+2\star\left(d\Phi\wedge \xi\right)\\
&=\frac{2}{\sqrt{-g}}\left(-\Phi\nabla_{[1}\xi_{0]}+2(\nabla_{[1}\Phi)\xi_{0]}\right).
\end{split}\label{QJT}
\end{equation}
The pre-symplectic form integrated on a Cauchy slice from a point $p_0$ on the surface $\sigma=0$ to a point $p_1=q(p_0)$ on the surface $\hat{\sigma}=2\pi$ is therefore given by
\be\label{sympresult}
\wt{\Omega}=s(q)\,\delta\left(Q_\xi|_{p_1}-Q_\xi|_{p_0}\right),
\ee
where
\begin{equation}
    s(q) = \begin{cases}
        +1 &\text{if }p_1\equiv q(p_0)\text{ lies to the right of }p_0,\\
        -1 &\text{if }p_1\equiv q(p_0)\text{ lies to the left of }p_0.\\
    \end{cases}
    \label{s}
\end{equation}

To simplify the evaluation of the Noether charge, it is useful to observe that in our expression \eqref{qvec} for the vector $\xi^{\hat{\alpha}}$ the variation $\delta f^\alpha$ is automatically tangent to the $dS_2$ surface in the embedding space, so we can change variables in the sum over $\alpha$ from $(\tau,\sigma)$ to $X^i$.  We thus have
\be
\xi^{\hat{\alpha}}(\hat{\tau},\hat{\sigma})=\frac{\partial(f^{-1})^{\hat{\alpha}}}{\partial X^i}\left(\delta e^\Q\right)^{i}_{\phantom{i}j}\hat{X}^j.
\ee
Defining
\be
\omega^{i}_{\phantom{i}j}=\left(e^{-\Q}\delta e^{\Q}\right)^i_{\phantom{i}j},
\ee
we then have
\begin{align}\nonumber
\Phi&=\phi\, Q_i\left(e^\Q\right)^i_{\phantom{i}j} \hat{X}^j\\\nonumber
\xi_{\hat{\tau}}&=-\omega^0_{\phantom{0}j}\hat{X}^j\\
\xi_{\hat{\sigma}}&=\hat{X}^1\omega^2_{\phantom{2}j}\hat{X}^j-\hat{X}^2\omega^1_{\phantom{2}j}\hat{X}^j.
\end{align}
From here computing the Noether charge \eqref{QJT} is straightforward algebra (we used Mathematica); the result is
\be\label{noetherresult}
Q_\xi=-\phi \,\delta\left(Q_i Q^i\right)
\ee
and thus
\begin{equation}
    \Omegatilde= -s(q)\, \delta\left( \phi\,\delta (Q_iQ^i) \right) = -s(q)\,\delta\phi\wedge\delta (Q_iQ^i).
    \label{dphidq2}
\end{equation}
This has rank two, so the phase space is indeed two-dimensional.  We emphasize that away from the case of $n=0$ with $Q^i$ spacelike, $s(q)$ is a class function for the adjoint action, meaning that $s(hqh^{-1})=s(q)$ for all $h\in \wt{G}$, so this symplectic form is well-defined on the set of solutions quotiented by the adjoint map.  For the inverse map we instead have $s(q^{-1})=-s(q)$, but the inverse map also flips the  sign of $\phi$, since $V^i$ is unchanged, so this symplectic form is also well-defined after the quotient by the inverse map.  

\bfig
\includegraphics[height=3cm]{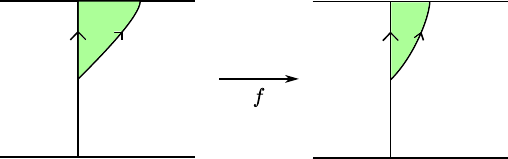}
\caption{The map to the standard gluing for a spacelike $n=0$ solution.}\label{n0glue2fig}
\efig
Returning now to the case where $n=0$ with $Q^i$ spacelike, our approach will instead be to take $f$ to do nothing near $\hat{\sigma}=0,\hat{\tau}\neq 0$, while near the image of $\hat{\sigma}=0$ under $q$ we have
\begin{equation}
\begin{split}
    f^\tau(\hat\tau,\hat\sigma) &= \arctan\left[ \left(e^{\mc{Q}}e^{-\Q_{ref}}\cdot \hat{X}\right)^0 \right], \\
    f^\sigma(\hat\tau,\hat\sigma) &= \arctan\left[ \frac{\left(e^{\mc{Q}}e^{-\Q_{ref}}\cdot \hat{X}\right)^2}{\left(e^{\mc{Q}}e^{-\Q_{ref}}\cdot \hat{X}\right)^1} \right],
\end{split}
\end{equation}
and
\begin{equation}
\begin{split}
    (f^{-1})^{\hat\tau}(\tau,\sigma) &= \arctan\left[ \left(e^{-\mc{Q}}e^{\Q_{ref}}\cdot X\right)^0 \right],\\
    (f^{-1})^{\hat\sigma}(\tau,\sigma) &= \arctan\left[ \frac{\left(e^{-\mc{Q}}e^{\Q_{ref}}\cdot X\right)^2}{\left(e^{-\mc{Q}}e^{\Q_{ref}}\cdot X\right)^1} \right].
\end{split}
\end{equation}
In other words, $f$ now maps the image of $\hat{\sigma}=0$ under $q$ to its image under some fixed spacelike transformation $q_{ref}$. See figure \ref{n0glue2fig} for an illustration. In our Noether charge calculation above, this causes the replacement $\omega \to e^{\Q_{ref}}\omega\, e^{-\Q_{ref}}$ in the expressions for $\xi_{\hat{\tau}}$ and $\xi_{\hat{\sigma}}$, and the dilaton becomes
\be
\Phi=\phi\, Q_i(e^{\Q}e^{-\Q_{ref}})^i_{\phantom{i}j} \hat{X}^j.
\ee
The Mathematica evaluation of the Noether charge now takes a bit longer, but the expression \eqref{noetherresult} holds without modification.  The pre-symplectic form is thus given by \eqref{sympresult}, just as before. There is, however, a new subtlety: when $n=0$ and $Q^i$ is spacelike, $s(q)$ is no longer a class function under the adjoint action!  Dealing with this is our next order of business.

\section{Group-theoretic description}
\label{sec:group-theory}
Let's now take stock of the situation.  In Section \ref{sec:sols}, we showed that solutions can be parameterized by an isometry $q\in\Gtilde$ (where $q$  can be represented by a pair $(Q^i,n)$ with $Q^i \in \mathcal R$ ---see (\ref{R})--- and $n\in\Z$) and a vector $V$ collinear with $Q$ associated to the dilaton profile. We also showed that, up to an issue for spacelike solutions with $n=0$, we can count each solution once by imposing the quotients \eqref{ids}.  In this section we will resolve the issue at $n=0$ by modifying our quotient in a subtle way, leading to a general group-theoretic construction of the phase space and symplectic form.

\subsection{Phase space}
\label{sec:gt-ps}
Let's first try to give the identifications \eqref{ids} a group-theoretic interpretation.  The quotient $(q,V)\sim (hqh^{-1},hV)$, from a group-theoretic point of view, is just the quotient by the differential 
\begin{align}
\label{Adstar}
      Ad_* ~:~  (q, V) \mapsto (h q h^{-1}, hV). 
\end{align} 
of the adjoint map
\be\label{}
Ad_h:q\mapsto hqh^{-1}.
\ee
On the other hand the quotient $(q,V)\sim (q^{-1},V)$ is not so natural: if we define a map
\be
Inv:q\mapsto q^{-1}
\ee
then its differential would act on $T\Gtilde$ as
\be
Inv_*:(q,V)\mapsto(q^{-1},-V),
\ee
which is not what we found.  Our expression \eqref{dphidq2} for the symplectic form suggests a way out.  Namely, we have
\begin{equation}
\begin{split}
\wt{\Omega}&=-s(q)\delta\phi\wedge\delta(Q_iQ^i)\\
&=-2\delta(s(q)\phi Q_i)\wedge \delta Q^i\\
&=-2\delta(s(q)V_i)\wedge \delta Q^i,
\end{split}
\end{equation}
so a more natural variable to consider than $V^i$ is 
\be\label{dilPtilde}
P_i=-2s(q)V_i.
\ee
Since $s(q^{-1})=-s(q)$, if we write the inverse action on $(q,P)$ instead of $(q,V)$ we have
\be
(q,P)\sim (q^{-1},-P),
\ee
just as we would want for the codifferential $Inv^*:T^*\wt{G}\to T^*\wt{G}$ of the inverse map:
\be
Inv^*:(q,P)\mapsto(q^{-1},-P).
\ee
Inspired by this success we can also try rewriting the codifferential adjoint action in terms of $P$:
\be
Ad_h^*:(q,P)\mapsto(hqh^{-1},Ph^{-1}).
\ee
Is this equivalent to what we had for $V$?  Almost!  To the extent that $s(q)$ is a class function, meaning it assigns the same sign to all members of each conjugacy class, then we can just multiply $V$ by $s(q)$ without changing the identification.  However, this property fails precisely when $n=0$ and $Q^i$ is spacelike!  A spacelike $Q^i$ corresponds to a boost in some direction, and by conjugating a boost by a rotation by $
\pi$ we can invert it. Thus, a rotation by $\pi$ in our old identification \eqref{ids} would have identified $(q,V)$ and $(q^{-1},-V)$, on top of the identification of $(q,V)$ with $(q^{-1},V)$ from the second line of \eqref{ids}, while our new identification \textit{only} identifies $(q,V)$ with $(q^{-1},V)$. We thus get back our missing solutions at $n=0$ with $Q^i$ spacelike!  This also solves a related problem with our expression for the symplectic form: if points in phase space are conjugacy classes, why should the non-class function $s(q)$ appear in the symplectic form?  With our new identification, it doesn't:
\be\label{Omegafinal}
\wt{\Omega}=\delta P_i \wedge \delta Q^i.  
\ee
We thus have arrive at the following proposal for the phase space:
\begin{equation}
    \faktor{T^\ast(\Gtilde\backslash \id)}{\{\sim Ad^\ast,\,\sim Inv^\ast\}} \quad\text{with the constraint}\quad Pq=P \quad\text{for all}\quad (q,P)\in T^\ast(\Gtilde\backslash \id).
    \label{gt-ps}
\end{equation}

In principle we could already declare victory here, but in most standard quantization methods one assumes that the classical phase space is a cotangent bundle, so it is interesting for us to understand to what extent that is the case here.  What we will show is that indeed the phase space (\ref{gt-ps}) is isomorphic to the cotangent bundle of the quotient $(\Gtilde \backslash \id)/\{\sim Ad^*, \sim Inv^*\}$, provided that we are careful in defining what we mean by the cotangent space at the singular points. The elements of the phase space (\ref{gt-ps}) are orbits
\begin{equation}
    [(q,P)] = \{ (h^{-1}q^\kappa h,\kappa P h)\: \vert \:h\in\Gtilde ,\,\kappa \in \{ -1,1\}\}
\end{equation}
restricted to pairs $(q,P)$ that satisfy $P q=P$. The  quotient $(\Gtilde\backslash\id)/\{\sim Ad,\sim Inv\}$ becomes the space of identification classes or ``base'' orbits
\begin{equation}
    [q] = \{ h^{-1}q^\kappa h\: \vert \:h\in\Gtilde ,\,\kappa\in\{-1,1\}\}.
    \label{qorb}
\end{equation}
Let us group orbits $[(q,P)]$ if they live at the same base orbit $[q]$, that is, if they can be expressed with the same base representative $q$. Two orbits $[(q,P)]$ and $[(q',P')]$ live at the same base orbit $[q]$ if and only if $q'\in[q]$, since then
\begin{equation}
    q'=h^{-1}q^{\kappa} h
\end{equation}
for some $h\in\Gtilde$, $\kappa\in\{-1,1\}$, and therefore
\begin{equation}
    [(q',P')]=[(q,\kappa P'h^{-1})].
\end{equation}
Moreover, $[(q,P)]$ and $[(q,P')]$ are equal if and only if
\begin{equation}
    P' = \kappa Ph
    \label{redundancy}
\end{equation}
for some $(h,\kappa)\in\Gamma_q$, where $\Gamma_q$ is the stabilizer of $q$:
\begin{equation}
    \Gamma_q \equiv \{ (h,\kappa)\in \Gtilde\times \{-1,1\}\:|\: h^{-1}q^\kappa h=q \}.
\end{equation}
Thus we can define
\begin{equation}
    [P]_q = \{\kappa P h\:|\:(h,\kappa)\in\Gamma_q,\:Pq=P\}
    \label{Porb}
\end{equation}
as an orbit of cotangent vectors at the base orbit $[q]$ (note that we have included the constraint $Pq=P$ in this definition).  The cotangent orbits $[P]_q$ can be constructed as the elements of the following quotient space:
\begin{equation}
    T^\ast_q(\Gtilde\backslash\id)\,/\,T^\ast_q[q]\,/\,\Gamma_q,
\label{tan}
\end{equation}
where the quotient by $\Gamma_q$ takes care of the redundancy (\ref{redundancy}) and the quotient by $T^\ast_q[q]$ imposes $Pq=P$, because it trivializes all $P$ in $T^\ast_q(\Gtilde\backslash\id)$ for which $Pq\neq P$:
\begin{equation}
\begin{split}
    T^\ast_q[q] &= \{P\in T^\ast_q(\Gtilde\backslash\id)\:|\: e^{-P_iT^i}q e^{P_iT^i}\neq q\}\\
    &= \{P\in T^\ast_q(\Gtilde\backslash\id)\:|\: q^{-1}e^{-P_iT^i}q\neq e^{-P_iT^i}\}\\
    &= \{P\in T^\ast_q(\Gtilde\backslash\id)\:|\: e^{-(P q)_iT^i}\neq e^{-P_iT^i}\}\\
    &= \{P\in T^\ast_q(\Gtilde\backslash\id)\:|\:P q\neq P\}.
\end{split}
\end{equation}
We have thus written the cotangent space at each base point $[q]$ in the convenient form (\ref{tan}) (with $q$ any representative of $[q]$). This expression is convenient because, inspired by pointwise calculations of the (co)tangent spaces of a quotient space, (\ref{tan}) seems to be the natural definition of the cotangent space at points $[q]$ in the quotient space
\begin{equation}\label{phase2}
    \faktor{\Gtilde\backslash\id}{\sim Ad,\sim Inv}\:\:,
\end{equation}
and note that this definition is well defined even at the singular points.\footnote{The expression for the tangent spaces given in (\ref{tan}) can be viewed as a generalized expression that combines the standard expressions for the tangent spaces of orbifolds by finite groups (where one must quotient by $\Gamma_q$; see, e.g., \cite{adem2007orbifolds}) and the tangent spaces for quotients by continuous Lie group actions (where one must quotient by $T^\ast_q[q]$ \cite{lee2012introduction}).} We have thus constructed a bijection between (\ref{gt-ps}) and the cotangent bundle
\begin{equation}
    T^\ast\left(\faktor{\Gtilde\backslash\id}{\sim Ad,\sim Inv}\right),
    \label{cotanbundle}
\end{equation}
so this is an equivalent way to write the phase space.

\subsection{Symplectic form}
In this section, we complete our group-theoretic story by showing that the symplectic form we calculated in Section 3 (and revisited in (\ref{Omegafinal})) turns out to be the canonically-defined\footnote{That is, canonically-defined with respect to the projection of the cotangent bundle to the base.} symplectic form on the group-theoretic formulation of the phase space that we constructed in the previous subsection. In particular, we will show (first for (\ref{cotanbundle}) and then, less abstractly but more tediously---and also more rigorously---, again for the equivalent expression (\ref{gt-ps})) that the symplectic form is the functional exterior derivative of the Liouville 1-form $\theta$ (see, e.g., \cite{AbrahamMarsden}),
	\begin{align}
		 \Omega = \delta \theta. 
	\end{align}
Although the computation of the above symplectic form for (\ref{cotanbundle}) will be straightforward, it is important to note that this computation assumes that the underlying geometry is a smooth symplectic manifold. However, since the cotangent bundle (\ref{cotanbundle}) is not a smooth manifold due to the presence of quotient singularities, this means that our canonical computation of the symplectic form is, in principle, not rigorous at singular points.\footnote{We could try to be more rigorous, for instance, by excising points of $\Gtilde\backslash\id$ where it fails to be a smooth manifold, such as singular points due to central elements $T^n \in \widetilde G$, as well as points where $\Gtilde\backslash\id$ is non-Hausdorff. This would enable us to obtain an expression for the symplectic form on a dense open set of $\Gtilde\backslash\id$.} It is to address this potential complication that we also compute, as a cross-check, the symplectic form on (\ref{gt-ps}); specifically, we compute the canonical symplectic form on $T^* \widetilde{G}$ (which is adequately smooth), and subsequently impose constraints that gauge the symmetries $Ad$ and $Inv$.\footnote{We ignore the identity element $\id \in \Gtilde$.} We show that, after restricting the symplectic form on $T^* \widetilde G$ to the subspace where these constraints are satisfied, and either fixing or projecting out the gauge symmetries, the resulting symplectic form agrees with the canonical symplectic form on (\ref{cotanbundle}).

Before delving into the details of the computation, we briefly review some relevant facts about symplectic geometry. Given any smooth manifold $M$ with cotangent bundle $\pi: T^*M \rightarrow M$, there is a canonical 1-form on $T^*M$ known as the tautological or Liouville 1-form (see, e.g., \cite{AbrahamMarsden}), often denoted $\theta$, which is defined as follows. Let $(q,p)$ denote a point in $T^*M$, where $q \in M$ and $p \in T^*_qM$. Then, given any tangent vector $X \in T_{(q,p)}(T^*M)$, the Liouville 1-form $\theta$ evaluated at the point $(q,p)$ is defined by
	\begin{align}
		\theta_{(q,p)}(X) = p(\D \pi_{(q,p)}(X)).
	\end{align}
Note that the differential $\D\pi$ of the projection map $\pi$ projects a tangent vector in $T_{(q,p)}(T^*M) \cong T_qM \times T_p(T_q^*M)$ to a tangent vector in $T_{q}M$. In local coordinates $(q^i, p_i)$, $\theta$ can be expressed as $\theta = p_i \D q^i$.
Then, up to a choice of overall sign, the symplectic form is canonically defined to be the exterior derivative of $\theta$, namely $\Omega = \D\theta$.

In our case, for (\ref{cotanbundle}), we take $M = (\Gtilde \backslash \id)/\{\sim Ad,\sim Inv\}$. An element of $T^\ast M$ can be written as $([q,P]) = ([q],[P]_q)$, and therefore
\begin{equation}
    \theta = [P_i\delta Q^i]_q\left( \left[ \delta Q^j\frac{\delta}{\delta Q^j} \right] \right),
\end{equation}
which takes the same value at every element in the orbit. Therefore, we can evaluate $\theta$ at one orbit representative, to obtain 
\begin{equation}
    \theta = P_i\,\delta Q^i.
\end{equation}
Then the symplectic form is
\begin{equation}
\label{formalOmega}
    \Omega = \delta P_i\wedge\delta Q^i,
\end{equation}
which matches our gravitational result, (\ref{dphidq2}), with $q$ labeling the spacetime quotient and the dilaton given by (\ref{dilPtilde}), consistent with Section \ref{sec:gt-ps}.

To verify that (\ref{formalOmega}) is correct even at singular points of the phase space, we now redo the calculation for (\ref{gt-ps}) instead, as a cross-check: we compute the symplectic form on $T^* \widetilde G$ and then introduce first- and second-class constraints to quotient by the group actions $Ad^*$ and $Inv^*$. First, we review some facts about Hamiltonian systems on Lie groups (see, e.g., \cite{alekseevsky1994poisson} and references therein). As before, we consider a cotangent bundle $\pi : T^* M \to M$, but this time we specialize to the case that $M$ is a Lie group $G$ with Lie algebra $\mathfrak{g}$, so that our (pre-)phase space is $T^* G$, which we may think of as consisting of elements $(q,P)$ with $q \in G, P \in \mathfrak{g}$. Let us write a vector in $ T_{(q,P)}(T^* \Gtilde)$ as $ (Q,P')$. Then, the Liouville 1-form $\theta$ on $T^* G$ is characterized by the property $\theta_{(q,P)}(Q,P') = P(Q)$. The Maurer-Cartan 1-form \cite{sternberg1999lectures} $\alpha$ on $\Gtilde$ naturally induces a Lie-algebra-valued 1-form on $T^* \Gtilde$, and thus we will use $\alpha$ to determine $\theta$ and its corresponding coordinate expression. When $G$ is a matrix group, then given a group element $q \in G$ one can write the Maurer-Cartan form explicitly as
	\begin{align}
		 q^{-1} \D q= \alpha^i T_i,
	\end{align}
where $T_i$ are basis elements of $\mathfrak g$. Introducing a dual basis $\{ t^i\}$ of $ \mathfrak g^*$, and writing the momenta as $P = P_i T^i$, the 1-form $\theta$ can be expressed as 
	\begin{align}
		\theta = \langle P, q^{-1} \D q \rangle = P_i \alpha^i,
	\end{align}	
where $\langle , \rangle$ is the Killing form, for which $\langle T^i, T_j \rangle = \delta^i_j$. Computing the exterior derivative and using the Maurer-Cartan equation $d\alpha = - \frac{1}{2} [ \alpha, \alpha]$ (which, using $[T_i,T_j] = \tensor{f}{^k_{ij}} T_k$ reads in components $d\alpha^i = - \frac{1}{2} f^i_{jk} \alpha^j \wedge \alpha^k$), we obtain
\begin{align}
	\Omega &= \D P_i \wedge \alpha^i  - \frac{1}{2} \tensor{f}{^i_{jk}} P_i \alpha^j \wedge \alpha^k. 
\end{align}
Although in our case $\widetilde G$ is not a matrix group, we may regard $\Gtilde$ as a finite central extension of $SO^+(2,1)$ with identical Lie algebra, and thus it suffices to use representatives of $SO^+(2,1)$ to compute the Maurer-Cartan form. Thus, we may work with elements $q_0 =e^{Q^i T_i} \in SO^+(2,1)$.

As a warmup exercise, let us use the fact that $SO^+(2,1) \cong \mathbb PSL(2,\mathbb R)$ and work in the spinor representation. We choose our generators $T_i$ in this representation to satisfy 
    \begin{align}
        T_i T_j = \eta_{i j} I + \tensor{\epsilon}{^k_i_j} T_k.
    \end{align}
(Although we later repeat the same calculation in the vector representation and obtain an identical result, the intermediate steps in the spinor representation are less complicated and thus much more illuminating.) It is straightforward to show that, in this representation,\footnote{In this section only we write $Q^2\equiv Q^iQ_i$ for simplicity.}
	\begin{align}
        \begin{split}
		q_0 &= \cosh \sqrt{Q^2} I + \frac{\sinh\sqrt{Q^2}}{\sqrt{Q^2}} Q^i T_i,
        \end{split}
	\end{align}
where $Q^2 = Q_i Q^i = -(Q^0)^2 + (Q^1)^2 + (Q^2)^2$. The variation of $q_0$ is given by
        \begin{align}
        \begin{split}
            \delta q_0 &= \sinh \sqrt{Q^2}  \:\delta \sqrt{Q^2} I + \frac{\cosh \sqrt{Q^2}}{\sqrt{Q^2}} \,\delta \sqrt{Q^2} Q^i T_i \\
            &+\frac{\sinh \sqrt{Q^2}}{\sqrt{Q^2}}\left(- \frac{ \delta \sqrt{Q^2}}{\sqrt{Q^2}} Q^i  +  \delta Q^i \right) T_i. 
        \end{split}
        \end{align}
    Using the above variation, the Maurer-Cartan 1-form can be written as 
        \begin{align}
        \begin{split}
        \alpha &\equiv \tensor{\alpha}{_j^k} \delta Q^j T_k = (A  \tensor{\epsilon}{^k_i_j}Q^i \delta Q^j + B \delta \sqrt{Q^2} Q^k + C \delta Q^k ) T_k ,
        \end{split}
        \end{align}
where
    \begin{align}
    \begin{split}
    \label{Qfns}
        A &= - \frac{\sinh^2 \sqrt{Q^2}}{\sqrt{Q^2}} \\
        B &= \frac{1}{\sqrt{Q^2}} - \frac{\sinh \sqrt{Q^2} \cosh \sqrt{Q^2}}{Q^2}\\
        C&= \frac{\sinh \sqrt{Q^2} \cosh \sqrt{Q^2}}{\sqrt{Q^2}}.
    \end{split}
    \end{align} 
    We dramatically simplify the computation at this stage by introducing first class constraints \cite{henneaux1992quantization} that will gauge the continuous symmetry $Ad^\ast$, namely the adjoint action of $SO^+(2,1)$, namely	\begin{align}
		\begin{split}
            \label{constraint}
			J_i = 0,~~~~J_i \equiv \epsilon_{ijk} Q^j P^k.
		\end{split}
	\end{align}
The constraint functions $J_i$ are proportional to the infinitesimal generators of the $SO^+(2,1)$ action on phase space, and we can think of the constraints as resulting from adding Lagrange multiplier terms to the Hamiltonian (which happens to be zero) that restrict the system to the subspace of phase space where the generators of $SO^+(2,1)$ are trivial. Notice that these  constraints force $Q^i, P^i$ to be collinear, which in turn imposes the stabilizer constraints $P q_0= q_0$. This implies that we may interpret the stabilizer constraints as first class constraints. 

Proceeding with the calculation of the symplectic form, we take the momentum covector $P_i$ to be collinear with $Q_i$, which ensures that the above constraints are satisfied. We write the proportionality factor in a way that makes contact with our findings from the previous subsection:
    \begin{align}
        P_i = -2 s(q)\phi\, Q_i.
    \end{align} 
Then, writing $\tensor{A}{^i_j} = A Q^k \tensor{\epsilon}{^i_k_j}$ for convenience, and using\footnote{Notation: Using the mostly $+$ convention for a $D$-dimensional spacetime, we use the identity $\epsilon_{\mu_1 \cdots \mu_n \nu_1 \cdots \nu_p} \epsilon^{\mu_1 \cdots \mu_n \rho_1 \cdots \rho_p} =  n! \delta^{\rho_1 \cdots \rho_p}_{\nu_1 \cdots \nu_p}$ where $p = D-n$ and the generalized Kronecker delta is defined by $\delta^{\rho_1 \cdots \rho_p}_{\nu_1 \cdots \nu_p} = \delta_{\nu_1}^{[\rho_1} \delta_{\nu_2}^{\rho_2} \cdots \delta_{\nu_p}^{\rho_p]}$. Here, (anti)symmetrization is accompanied by an overall normalization of $1/p!$ (e.g., $T_{[\alpha \beta]} = (T_{\alpha \beta} - T_{\beta \alpha})/2$).} $Q_i \alpha^i = \delta Q^2/2 $ and $\delta Q_i \wedge \alpha^i = A_{ij} \delta Q^i \wedge \delta Q^j$,
we find that the symplectic form is given by 
        \begin{align}
        \begin{split}
        \label{groupOmega} 
          \Omega& =  -s(q)\, \delta \phi \wedge \delta Q^2+    s(q)\, \phi\, (2A+ A^2 Q^2+ C^2 )\, \epsilon_{k i j}\, Q^k   \delta Q^i \wedge \delta Q^j,
        \end{split}
        \end{align}
with $A,C$ given in (\ref{Qfns}). Notice that the first term in the last line of the above equation agrees precisely with the symplectic form in (\ref{dphidq2}), while the second term vanishes when the gauge symmetry is either projected out or fixed, as this implies $\delta n^i = 0$. Hence, we find that 
    \begin{align}
        \Omega =  -s(q)\, \delta \phi \wedge \delta Q^2 = \delta P_i \wedge \delta Q^i. 
    \end{align}

To be completely consistent with our parametrization of the phase space, we repeat the above computation in the case that $q_0$ is in the vector representation of $SO^+(2,1)$, using the algebraic properties described in Appendix \ref{app:psl}. The Maurer-Cartan form $\alpha$ is again straightforward to compute, but requires much more algebra and thus we do not reproduce the details of the computation here. Using a symbolic computing tool such as Mathematica, one can easily verify that $\alpha$ can be expressed entirely in the Lie algebra basis $\{T_i\}$, and that $Q^i \alpha_i = \delta Q^2/2$ as before. Thus, again, the symplectic form is  given symbolically by (\ref{groupOmega}), where the second term on the hand side is eliminated by projecting out or fixing the gauge symmetry. Finally, we may quotient out the discrete symmetry $Inv^\ast$ to remove the prefactor $s(q)$.

\section{Comments on quantization}
\label{sec:quant}

Having constructed the classical phase space of JT gravity with positive cosmological constant, the next natural step is to quantize it.  Based on our two equivalent constructions \eqref{gt-ps} and \eqref{phase2} of the phase space, there are two natural quantization procedures to consider.  Starting from \eqref{phase2} is conceptually simpler: we have a fully gauge-fixed cotangent bundle, so we can simply define the Hilbert space to be the space of square-integrable functions
\be
\mathcal{H}=L_2\left(\faktor{\Gtilde}{\sim Ad,\sim Inv}\right),
\ee
with a rule that the wave function must vanish at the identity class.  This seems nice and clean, except that this is a rather singular space and so some mathematical care is likely needed to make sense of it.  If we instead start from \eqref{gt-ps} then the starting point is nicer: we construct a pre-Hilbert space
\be
\wt{\mathcal{H}}=L_2(\wt{G}),
\ee
where, since $\wt{G}$ is a nice smooth Lie group, there is no subtlety, and again we demand that the wave function vanish at the identity. But then, to get to the physical Hilbert space, we need to impose three constraints: we restrict to states which are invariant under $Ad^*$ and $Inv^*$, and impose the condition that $Pq=P$.  Our expectation is that these two quantization methods should lead to the same results, but we have not yet undertaken a systematic study.  

\paragraph{Acknowledgments} We thank Chris Akers, Sergio Hern\'andez-Cuenca, David Kaiser, David Kolchmeyer, Adam Levine, Jennie Traschen, Nico Vald\'es-Meller, and (especially) Jon Sorce for helpful discussions. Portions of this work were conducted in MIT’s Center for Theoretical Physics and supported in part by the U. S. Department of Energy under Contract No.~DE-SC0012567. EAM is supported by a fellowship from the MIT Department of Physics. This research was also supported in part by grant NSF PHY-2309135 to the Kavli Institute for Theoretical Physics (KITP). PJ is supported by the Johns Hopkins University Provost's Postdoctoral Fellowship, and in part by the Simons Collaboration on Global Categorical Symmetries and grant NSF PHY-2112699. DH is supported by the Packard Foundation as a Packard Fellow, the Air Force Office of Scientific Research under the award number FA9550-19-1-0360, the US Department of Energy under grants DE-SC0012567 and DE-SC0020360, and the MIT department of physics.

\appendix

\section{Vector parameterization of the Lorentz group in 2+1 dimensions}
\label{app:psl}

In this appendix we give the argument that elements of $SO(2,1)$ are in one-to-one correspondence with the region $\mathcal{R}$ defined by equation \eqref{Rdef}.  Every element $q$ of $G=SO(2,1)$ can be generated by exponentiation of an element $\mathcal{Q}=Q^iT_i$ of its Lie algebra $\so(2,1)$, $q=e^\mathcal{Q}$, where $Q^i$ is a real 3-vector and $T_i$ are the Lie algebra generators.\footnote{This is not obvious, as for general noncompact Lie groups the exponential map is not surjective, but it is true.} We will take the generators to be:
\begin{equation}
\label{vecgenerators}
    T_0=\begin{pmatrix}
        \,0&0&0\,\\\,0&0&-1\,\\\,0&1&0\,
        \end{pmatrix},\quad T_1=\begin{pmatrix}
        \,0&0&-1\,\\\,0&0&0\,\\\,-1&0&0\,
    \end{pmatrix},\quad T_2=\begin{pmatrix}
        \,0&1&0\,\\\,1&0&0\,\\\,0&0&0\,
    \end{pmatrix}.
\end{equation}
Note that the Lie algebra endows the space of vectors $Q^i$ with a metric $\mathcal{G}$, induced by the Killing form, namely
\begin{equation}
    \mathcal{G}(Q_1,Q_2) \equiv \frac{1}{2}\,\text{Tr}(\mathcal{Q}_1\cdot \mathcal{Q}_2),
    \label{metric}
\end{equation}
where $\mathcal{Q}_1=(Q_1)^iT_i,\, \mathcal{Q}_2=(Q_2)^iT_i$, and we are free to choose the normalization factor $1/2$, so that the square of $Q^i$ is
\begin{equation}
    \mathcal{G}(Q,Q)=Q^iQ_i=-(Q^0)^2+(Q^1)^2+(Q^3)^2,
\end{equation}
and thus $Q^i$ are vectors in Minkowski space. We want to use the vectors $Q^i$ as coordinates on the manifold $G$, to label points $q$. However, multiple $\mathcal{Q}=Q^iT_i$ generate the same $q$. We must quotient the space of $Q^i$ so that the fundamental domain is isomorphic to $G$ as a manifold, and therefore $Q^i$ are good coordinates on $G$.

First note that a simple timelike, spacelike, and null $Q^i$ generate
\begin{equation}
\begin{split}
    e^{Q^0T_0}=&\begin{pmatrix}
        1&0&0\\ 0&\cos Q^0&-\sin Q^0\\0&\sin Q^0 &\cos Q^0
    \end{pmatrix},\:e^{Q^2T_2} = \begin{pmatrix}
        \cosh Q^2 &\sinh Q^2&0\\ \sinh Q^2&\cosh Q^2&1\\0&0&1
    \end{pmatrix},\\
    &e^{Q^0(T_0-T_1)}=\begin{pmatrix}
        1+\frac{(Q^0)^2}{2} & \frac{(Q^0)^2}{2} & Q^0\\ -\frac{(Q^0)^2}{2} & 1-\frac{(Q^0)^2}{2} & -Q^0\\ Q^0&Q^0&1
    \end{pmatrix},
\end{split}
\end{equation}
which are a rotation, a boost, and a nontrivial combination of rotations and boosts. Clearly $Q^i=(Q^0,0,0)$ and $Q^i=(Q^0\pm 2\pi n,0,0)$ for any $n\in \Z$ generate the same $q$. More generally, for any $Q^i$,
\begin{equation}
    e^{\mathcal{Q}} = \id + \frac{\sinh\sqrt{Q^iQ_i}}{\sqrt{Q^iQ_i}}\mathcal{Q} + \frac{\cosh\sqrt{Q^iQ_i}-1}{Q^iQ_i}\mathcal{Q}^2,
\label{exp}
\end{equation}
where $\id$ is the $3\times 3$ identity matrix. Then
\begin{equation}\label{traceQ}
    \Tr e^{\mathcal{Q}} = 1 + 2\cosh\sqrt{Q^iQ_i},
\end{equation}
which satisfies $-1<\Tr e^{\mathcal{Q}}<3$, $\Tr e^{\mathcal{Q}}>3$ and $\Tr e^{\mathcal{Q}}=3$ when $Q^i$ is timelike, spacelike or null, respectively.  Moreover we have
\be\label{propref}
\frac{1}{2}\Tr \left(T_j e^\Q\right)=\frac{\sinh \sqrt{Q^iQ_i}}{\sqrt{Q^iQ_i}}Q_j.
\ee
Thus two vectors $Q^i$ and $Q^{i\prime}$ which exponentiate to the same group element must be proportional to each other.  In the null case the coefficient of proportionality is $\frac{\sinh 0}{0}=1$, so they must be equal.  In the spacelike case they must lead to the same trace of $e^\Q$, which by \eqref{traceQ} means they must have the same length.  From \eqref{propref} we then have
\be
\frac{\sinh \sqrt{Q^{i\prime}Q_i'}}{\sqrt{Q^{i\prime}Q_i'}}Q_i'=\frac{\sinh \sqrt{Q^iQ_i}}{\sqrt{Q^iQ_i}}Q_i'=\frac{\sinh \sqrt{Q^iQ_i}}{\sqrt{Q^iQ_i}}Q_i,
\ee
so again we must have $Q^{i\prime}=Q^i$.  In the timelike case, however, \eqref{traceQ} only tells us that
\be
\cos (i\sqrt{Q^{i\prime}Q_i'})=\cos (i\sqrt{Q^{i}Q_i}),
\ee
which means that must either have
\be\label{norm1}
i\sqrt{Q^{i\prime}Q_i'}=i\sqrt{Q^{i}Q_i}+2\pi n
\ee
or else
\be\label{norm2}
i\sqrt{Q^{i\prime}Q_i'}=-i\sqrt{Q^{i}Q_i}+2\pi m.
\ee
From \eqref{propref} we therefore see that given $Q^i$ we can construct a vector
\be
Q^{i\prime}=\frac{i\sqrt{Q^iQ_i}+2\pi n}{i\sqrt{Q^i Q_i}}Q^i
\ee
which exponentiates to the same group element as $Q^i$.  By choosing $n$ appropriately, we can therefore arrange for $Q^{i\prime}$ to obey
\be
Q^{i\prime}Q_i'\geq \pi^2.
\ee
When this inequality is not saturated, any other vector which exponentiates to the same group element ---and whose norm is thus related to that of $Q^i$ by \eqref{norm1} or \eqref{norm2}--- cannot satisfy this condition.  When the inequality is saturated, $Q^i$ and $-Q^i$ both exponentiate to the same group element, so we should choose only one of them. Without loss of generality, we choose the future-pointing one. This completes the argument that the region $\mathcal{R}$ defined by equation \eqref{Rdef} counts each element of $SO(2,1)$ exactly once.  

\bibliographystyle{jhep}
\bibliography{bibliography}
\end{document}